\documentclass[12pt]{article}
\usepackage{epsfig}
\begin{document}

\begin {center}
{\bf A helpful analogy between the covalent bond and Particle
Spectroscopy}
\vskip 5mm
{D.\ V.\ Bugg\footnote{email: d.bugg@rl.ac.uk}
\\ {\normalsize\it Queen Mary, University of London, London E1\,4NS,
UK} \\[3mm]}
\end {center}
\date{\today}

\begin{abstract}
\noindent
It is proposed that meson resonances are linear combinations of $q\bar
q$ and meson-meson (MM); baryon resonances are combinations of $qqq$
and meson-baryon (MB).
Mixing between these combinations arises via decays of confined states
to meson-meson or meson-baryon.
There is a precise analogy with the covalent bond in molecular physics;
it helps to visualise what is happening physically.
One eigenstate is lowered by the mixing;
the other moves up and normally increases in width.
Cusps arise at thresholds.
At sharp thresholds due to S-wave 2-particle decays, these
cusps play a conspicuous role in many sets of data.
The overall pattern of light mesons is consistent with nearly linear
Regge trajectories, hence $q\bar q$ components.
There is no obvious reason why this pattern should arise from
dynamically generated states without $q\bar q$ content.

\vskip 2mm

{\small PACS numbers: 12.39.Ki, 12.40.Yx, 13.25.Ft, 14.20.Gk}
\end{abstract}

\section {Introduction}
In the early days of Particle Physics, Chew, Goldberger and others
tried to account for resonances in terms of particle exchanges
\cite {Chew}.
This met with partial success.
For $\pi N$ elastic scattering, Donnachie and Hamilton \cite {Donnachie}
showed that exchanges of $N$, $\Delta(1232)$, $\rho$ and $\sigma$
provide long range attraction in $P_{33}$, $D_{13}$, $D_{15}$ and
$F_{15}$ partial waves where prominent resonances are observed.
Furthermore, these exchanges account for repulsive partial waves
$P_{13}$, ${P_{31}}$, $P_{11}$, $D_{33}$ and $D_{35}$.
However, meson exchanges failed to account for the $\rho$ meson.
This approach was therefore quickly overtaken by the quark model,
which has provided a semi-quantitative picture of most of the known
resonances.

However, there are cracks in this framework.
Firstly, it does not account for the $\sigma$ and $\kappa$ poles,
$a_0(980)$ and $f_0(980)$.
Secondly, there are many examples where resonances appear at or
close to sharp S-wave thresholds: e.g. $f_0(980)$ and $a_0(980)$
at the $K\bar K$ threshold, $f_2(1565)$ at the $\omega
\omega $ threshold, $X(3872)$ at the $\bar D_0 D_0^*$ threshold,
$S_{11}(1650)$ and $D_{13}(1700)$ close to the $\omega N$ threshold.

There is a straightforward explanation of how S-wave thresholds
attract these resonances \cite {Sync}.
Consider $f_0(980)$ as an example.
Its amplitude for $\pi \pi \to KK$ is given to first approximation
by the Flatt\' e formula
\cite {Flatte}: 
\begin {equation} A_{12} = T_{12}\sqrt {\rho _1 \rho _2} =
\frac {G_1G_2 \sqrt {\rho _1 \rho _2 }}
{M^2 - s - i[G_1^2 \rho _1(s) + G_2^2 \rho _2(s)]}
=\frac {N(s)}{D(s)}
\end {equation}
where phase space $\rho$ is factored out of $T$.
Here $G_i = g_i F_i(s)$ and $g_i$ are coupling constants, $F_i$ are
form factors.
Writing $D(s) = M^2 - s - i\Pi (s)$, a more exact form for $D(s)$ is
$M^2 - s - Re \, \Pi (s) - i\Pi (s)$, where
\begin {equation} Re \, \Pi _{KK}(s) =
\frac {1}{\pi} \rm {P} \int _{4m^2_K} ^\infty ds' \,
\frac {G^2_{KK}(s') \rho _{KK}(s')} {(s' - s)}.
\end {equation}

Fig. 1(a) illustrates the dispersive term ${\rm Re} \, \Pi (s)$ for
$f_0(980)$ using $F_{KK} = \exp (-3k^2_{KK})$, where $k_{KK}$ is $KK$
momentum in GeV/c.
${\rm Re} \, \Pi$ acts as an effective attraction.
Parameters of $f_0(980)$ are known.
Ref. \cite {Sync} gives tables of pole positions
when $M$, $G_1$ and $G_2$ are varied.
If $M$ is as low as  500 MeV, there is still a pole at
$806 - i78$ MeV;
for $M$ in the range 850--1100 MeV, there is a pole within 23 MeV
of the $KK$ threshold.
The moral is that a strong threshold can move a resonance a
surprisingly long way.
For $f_0(980)$, $G^2_{KK} \sim 0.7$ GeV$^2$.
\begin{figure}[htb]
\begin{center} \vskip -8mm
\epsfig{file=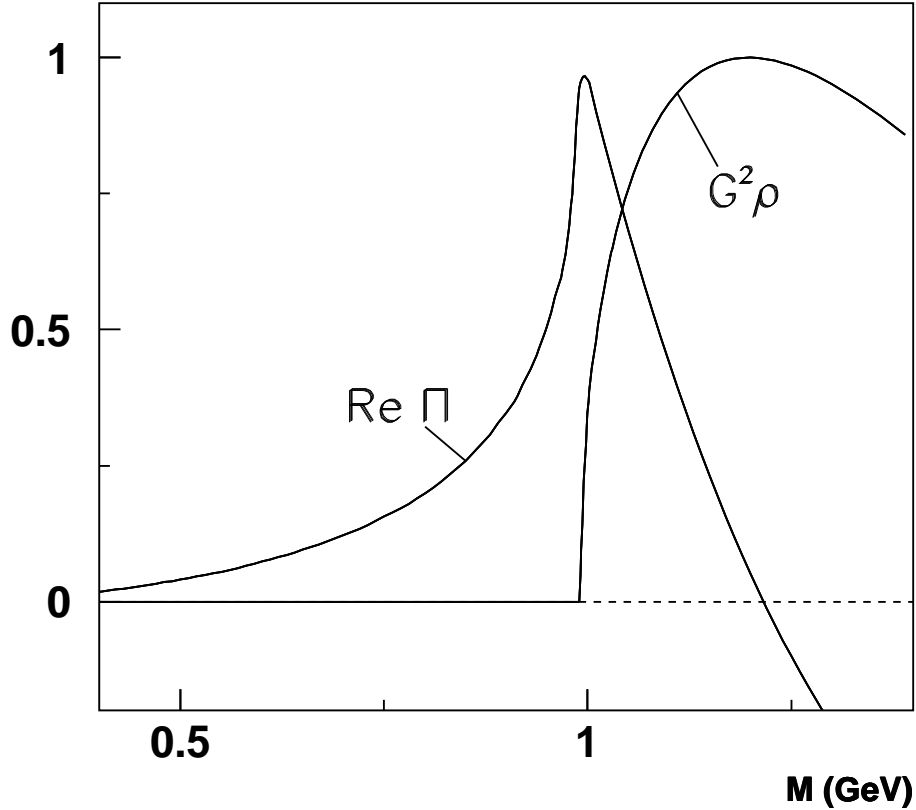,width=16cm}
\vskip -6mm
\caption
{(a) ${\rm Re} \, \Pi _{KK}(s)$ and $G^2_{KK}\rho _{KK}(s)$ for
$f_0(980)$, normalised to 1 at the peak of $G^2_{KK}\rho _{KK}$; (b)
the loop diagram for $f_0(980) \to K\bar K$.}
\end{center}
\end{figure}

It is important to realise that ${\rm Re} \, \Pi$ is not an `optional
extra'.
It is a rigorous consequence of analyticity for all $s$-channel
decay processes.
In principle these terms are required for all resonance decays.
It is then logical to include also $t$- and $u$-channel exchanges.

P-wave thresholds lead to broader effects because of the $k^3$
momentum dependence of phase space.
There are in principle contributions to $\rho (770)$ and $a_1(1260)$,
but in practice these effects may be accomodated by fitted
form factors describing decay widths.

The virtue of Eq. (2) is twofold.
Firstly, it is easily evaluated, secondly it illustrates graphically
the effect of the form factor $F$.
Re $\Pi (s)$ goes negative close to the peak of $G^2 \rho$ and
subsequently has a minimum at $\sim 1.7$ GeV; thereafter it slowly
rises to 0 as $M \to \infty$.

There are alternatives to evaluating Eq. (2).
The same result may be obtained by evaluating the loop diagram of
Fig. 1(b).
Secondly, solving the Bethe-Salpeter equation is equivalent to
evaluating all loop diagrams.
It is straightforward in principle to include $s$- and $t$-channel
exchanges in solving this equation.

Several authors have adopted a similar approach.
Jaffe \cite {Jaffe} gives the equations and discusses many
of the implications in Section 2C of his paper.
The Hamiltonian for a $q\bar q$ state decaying to meson-meson obeys
\begin {equation} H \Psi  = \left( \begin {array} {cc} H_{11} & V \\
V & H_{22} \end {array} \right) \Psi;
\end{equation}
$H_{11}$ describes short-range configurations;
$H_{22}$ refers to ingoing and outgoing states and should include
$t$- and $u$-channel exchanges;
$V$ accounts for the coupling between them due to $s$-channel decays.

Weinstein and Isgur pursued the connection between $q\bar q$,
$qq \bar q \bar q$ and meson components in their work on
$K\bar K$ molecules \cite {Weinstein}.
Van Beveren and Rupp \cite {eef} construct a model where
the short-range attraction in $H_{11}$ is approximated by a
harmonic oscillator potential, which couples at radius $R$ to
ingoing and outgoing waves corresponding to decay channels.
Despite the approximations, this gives valuable insight.
Their algebraic solution satisfies the Schr\" odinger equation
and is therefore fully analytic.
It includes effects of thresholds fully, although not yet
the $s$- and $t$-channel exchanges.

Barnes and Swanson \cite {Barnes} consider meson loops due to
pairs of $D$, $D^*$, $D_s$ and $D_s^*$ mesons, using the
$^3P_0$ model for decays.
For 1S, 1P and 2P charmonium states, they find that large mass shifts
due to these loops may be `hidden' in the valence quark model by a
change of parameters.
The important conclusion from their work is that
two-meson continuum components of charmonium states may be quite large,
with the result that the constituent quark model predicts masses which
are too high, particularly near the thresholds of opening channels.

Oset and collaborators demonstrate in a series of papers that some
resonances may be understood as `dynamically driven states'
\cite {Gamermann} \cite {Geng} \cite {Gonzalez} \cite {Geng2}
\cite {Molina} \cite {Sarkar} \cite {Molina2} \cite {Oset} due to
$s$, $t$ and $u$-channel exchanges.
They take the view that $\bar qq$ and $qqq$ components are not needed
at all in these cases.
This takes us full circle back to the approach tried by Chew.

How is it that Oset et al. are able to reproduce known resonances
(approximately) with meson exchanges alone?
They use S-wave form factors which are adjusted to get one predicted
resonance of each paper at the right mass.
Resulting amplitudes are strong.
The form factors may be mocking up short range $q\bar q$ components.
The importance of their work is the demonstration that
components derived from meson loops are large, and should
be taken into account.

On the other hand, the well known $J/\Psi$, $^3P_0(3415)$,
$^3P_1(3510)$, $^3P_2(3556)$, $\Psi (2S)$ and $\Psi (3770)$ are
interpreted naturally as $c\bar c$ states (with tiny admixtures of the
mesonic states to which they decay).
Therefore it is logical to include the $c\bar c$ component into all
other resonances unless there is a good reason why not.
One should not be deterred from invoking $q\bar q$ and $qqq$
components to get all resonances with their correct masses and widths.

The central premise of the present paper is that both $H_{11}$ and
$H_{22}$ play essential roles in all cases.
This is different from approaches where only one of the two
components in the Hamiltonian contributes, for example the approach
based on four-quark mesonic states.

An approach which has recently been popular is to suppose that
`molecules' are formed from $\bar q \bar q q q$ configurations
\cite {Maiani} \cite {Braaten}\cite {Hanhart}  \cite {Alvarez}
\cite {Martinez}.
The well known question over this approach is why so few tetraquarks
are observed.
Vijande et al. \cite {Vijande} throw light on this issue.
They study the stability of pure $c\bar c n \bar n$ and
$cc \bar n \bar n$ states in the absence of diquark interactions.
They find that all 12 $c\bar c n\bar n$ states with $J = 0$, 1 or 2
are unstable.
Their calculation points to the conclusion that such molecules are
rare unless either (i) there are attractive diquark interactions, or
(ii) coupling to meson-meson final states contributes, as proposed
here.

The layout of remaining sections is as follows.
Section 2 considers a useful analogy with the covalent bond in
chemistry.
This analogy helps visualise the main features of mixing.
Section 3 reviews approximations to be used in fitting data.
Section 4 considers $\sigma$, $\kappa$, $a_0(980)$ and $f_0(980)$.
Section 5 discusses firstly $X(3872)$ and concludes, as do several
authors, that data require a linear combination of $c\bar c$
and $\bar D D^*$.
Next it reviews the status of light mesons with $J^P=1^-$, $2^+$
and $0^+$.
The objective is to demonstrate that observed states lie
close to straight line trajectories when masses squared are
plotted against radial excitation number.
There are some significant deviations attributable to thresholds.
However, the overall picture is consistent with rather regularly
spaced $q\bar q$ states consistent with Regge trajectories.
Such trajectories are usually attributed to flux tubes joining
$q\bar q$ pairs and expanding as spin $J$ increases.
There is no obvious reason why dynamically generated
states should follow such a regular sequence.
Section 6 makes remarks on desirable experiments and Section 7
draws conclusions.

\section {A helpful analogy with the covalent bond in chemistry}
The wave function $\Psi$ of Eq. (3) is a linear combination of
$q\bar q$ (or $qqq$) and unconfined MM (or MB).
The key point is that two attractive components $H_{11}$ and $H_{22}$
lower the eigenstate via the mixing.
This is a purely quantum mechanical effect.
There is a direct analogy with the covalent bond in chemistry.
The solution of Eq. (3) is given by the Breit-Rabi equation:
\begin {equation}
E = (E_1 + E_2)/2 \pm \sqrt {(E_1 - E_2)^2 + |V|^2},
\end {equation}
where $E_1$ and $E_2$ are eigenvalues of separate $H_{11}$ and $H_{22}$.
One linear combination  is pulled down in energy.

In chemistry, $H$ is in principle known exactly.
The discussion of the hydrogen molecule (and more complex ones) is
given in many textbooks on Physical Chemistry, for example the one of
Atkins \cite {Atkins}.
Consider two hydrogen atoms labelled A and B, combining to make a
hydrogen molecule; $H_{11}$ and $H_{22}$ describe A and B.
The equation describing this pair is precisely the same as Eq. (3).
except that $H_{11}$ and $H_{22}$ have a different radial dependence
to the mesonic case.
Fig. 2(a) sketches contours of electron density for the lower
state of Eq. (4).
The effect of the mixing is that the electron density adjusts so
that the two electrons are somewhat concentrated between the two
ions.
In more detail, the wave-function for the atomic molecule
is expanded in terms of a complete set of atomic H orbitals.
For hydrogen, these are just the energy levels of a
single hydrogen atom.
For a carbon atom, they are replaced by wave functions allowing for
shielding of the Coulomb field of the nucleus by electrons in the
lowest atomic levels.
It is also necessary to anti-symmetrise wave functions.
Atkin gives algebraic forms which approximate the electronic
wave functions.
With modern computing techniques, the Variational method adjusts
the coefficients of the expansion in terms of excited states to
describe wave functions with great accuracy.
\begin{figure}[htb]
\begin{center} \vskip -20mm
\epsfig{file=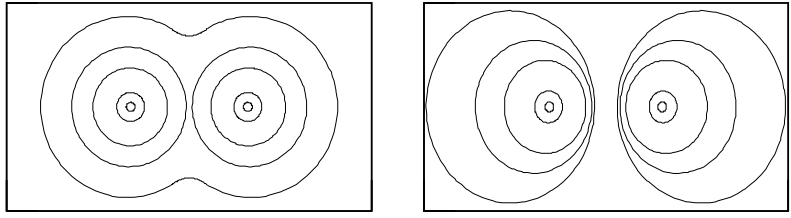,width=18cm}
\vskip -20mm
\caption
{Sketch of the electron density in the hydrogen molecule for
(a) lower and (b) upper states of Eq. (4).}
\end{center}
\end{figure}

Two key points are that (a) the extension of the wave function into the
overlap region lowers momenta of electrons, hence their kinetic energy,
(b) the whole system shrinks slightly and the binding of electrons to
both nuclei increases.

In the Particle Physics case, the procedure is
conceptually identical.
The mesonic wave function for the ground state is
sucked into the region of overlap, producing an attractive
interaction between $q\bar q$ and meson-meson.
This implies that $q\bar q$ will decay to final states where
the interaction is indeed attractive.
Fig. 2(b) shows the electron density for the upper state of Eq. (4).
In this case, electrons are repelled from the overlap region,
increasing their momenta and kinetic energies, hence the energy
eigenvalue.

The second effect is that the increased binding for the
lower state draws the quarks slightly down the Coulombic part of
the QCD potential, shrinking the radius of the state.
The converse happens for the upper state.
Calculation of the mixing requires the radial wave functions of
$q\bar q$ states, which can be evaluated
from the funnel potential.
A difficulty, however, is that the radial form of the mixing element
$V$ between $q\bar q$ and $MM$ parts of the wave function is unknown,
and has to be guessed.
The calculations of Oset et al. concern purely the mesonic part of the
wave function.

The conceptual analogy is simple: mixing between $q\bar q$ and
meson-meson will lower the eigenvalue of the lower of the two states
given by (4).
The Variational Principle governs this eigenvalue and the wave function.
The numerator of Eq. (2) is positive definite.
It is therefore unavoidable that ${\rm Re} \, \Pi(s)$  goes negative
at large $s$.
The Variational Principle will cut off the high mass tail of the
resonance, so as to minimise the repulsive part of ${\rm Re} \,
\Pi(s)$, hence reducing the resonance width.
A feature of light $q\bar q$ and $qqq$ resonances is that their widths
are roughly equal to the spacing between radial excitations.
It is likely that widths are restricted to this value by a feedback
mechanism which creates an orderly sequence of resonances.
A general feature of decays of high mass resonances is that S-wave
decays to low mass final states are weak.
Decays tend to be to high mass configurations with small
momenta.
High spin states generally decay with large angular momenta, where
the centrifugal barrier delays the opening of the threshold.

There is little evidence for decays to $I=2$ $\pi \pi$ pairs
or $I = 3/2~K\pi$, where interactions are repulsive \cite {kappa}.
The commonly observed SU(3) octets and decuplets are those whose decays
do {\it not} lead to such repulsive final states.
Higher representions such as {\bf 27}, {\bf 10} and $\bf {\bar {10}}$
do lead to such repulsive final states.
The natural interpretation is that repulsive final states actively
inhibit formation of representations higher than octets and decuplets.
The Variational Principle arranges that the configurations
produced are those where repulsive final states are suppressed.

\section {Approximations}
A difficulty at present is that the form factor used in Fig. 1 is not
known precisely.
The usual Flatt\' e formula, Eq. (1), serves as an approximate
fitting function where $M$ and $g^2$ are fitted empirically.
However, the cusp changes slope abruptly at the $KK$
threshold.
Analysis of data then requires a precise knowledge of experimental
resolution if the cusp is included in the fit.
This is illustrated for $a_0(980)$ in  Ref. \cite {a0980}, where
Crystal Barrel data on $\bar pp \to \eta \pi ^0 \pi ^0$ are fitted
including the cusp.
The mass resolution, 9.5 MeV,  is known accurately in this case, but
is large enough to smear out the cusp seriously.
The cusp plays a strong role if the energy resolution is as good as
1 MeV.
A further detail is  that separate thresholds for $K^+K^-$ and
$K^0 \bar K^0$ have not been used in Fig. 1 for simplicity; these two
thresholds may be taken into account straightforwardly and the
equations are given by Achasov and Shestakov \cite {Achasov}.

\subsection {Broad Thresholds}
Although broad thresholds may play a role in forming a resonance,
the dispersive term ${\rm Re} \, \Pi$ eventually has only a small
effect on the experimental line-shape in most cases.
There is, however, a crucial detail which has often been neglected
in fits to data.

Let us consider as an example $\pi \pi \to \rho \rho$.
The Breit-Wigner amplitude in this case is
\begin {equation}
f  = \frac {1}{k} \frac {M\sqrt {\Gamma _{\pi \pi}(s)}
\sqrt {\Gamma _{4\pi }(s)}}
{M^2 - s - {\rm Re} \, \Pi (s) - iM\Gamma _{total}(s)}.
\end {equation}
Here, $k$ is   the $\pi \pi$ centre of mass momentum (allowing for the
incident flux).
The $\pi \pi$ phase space is approximately constant.
However, it is essential to allow for the rapid $s$-dependence of
$\Gamma _{4\pi}$.
Many experimental analyses ignore this point and fit $|f|^2$ to a
Breit-Wigner resonance of constant width.

This is a critical point for many resonances in the mass range 1--2
GeV, where thresholds are opening.
Fig. 3(a) shows line-shapes of $f_0(1370) \to \pi \pi$ and $4\pi$ as
an example.
There is a large difference between them.
This is the source of the large spread in masses reported by the
Particle Data Group (PDG) for $f_0(1370)$ \cite {PDG}.
Anisovich et al. \cite {Anisovich} determine the K-matrix pole position
$1306 \pm 20$ MeV from a combined analysis of data on $\pi ^-p \to
\pi ^0 \pi ^0 n$ and $K\bar K n$, $\pi ^+\pi ^- \to \pi ^+\pi ^-$,
and Crystal Barrel data for $\bar pp$ at rest $\to 3\pi ^0$,
$\pi ^0 \eta \eta$, $\pi ^+\pi ^-\pi ^0$, $K^+K^-\pi ^0$,
$K_S^0 K_S^0\pi^0$, $K^+K_S^0\pi ^-$ and
$\bar pn \to \pi ^-\pi ^-\pi ^+$, $K_S^0K^-\pi ^0$ and
$K_S^0 K_S^0\pi ^-$; the last three determine P-state annihilation.
This analysis did not explicitly allow for ${\rm Re} \, \Pi(s)$
in the denominator.
That was taken into account in Ref. \cite {1370},
resulting in a mass of $1309 \pm 15$ MeV, in close agreement
with Anisovich et al.
The peak in $4\pi$, if judged from the mean of half-heights, is
1377 MeV, in good agreement with the determination of mass from
$4\pi$ data quoted by the PDG.
\begin{figure}[htb]
\begin{center} \vskip -10mm
\epsfig{file=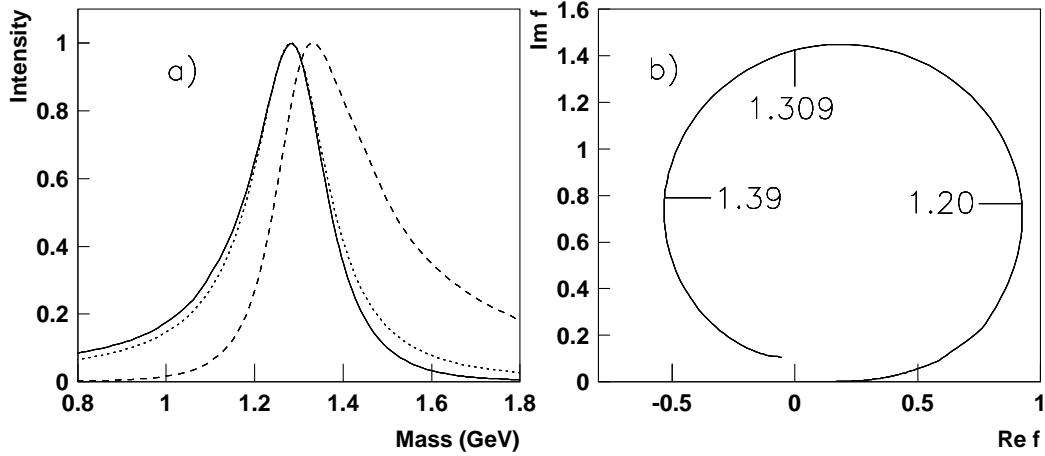,width=16cm}
\vskip -6mm
\caption
{(a) Line-shapes of $f_0(1370)$ for $\pi \pi \to \pi \pi$ (full),
a Breit-Wigner amplitude of constant width (dotted), and
$\pi \pi \to 4\pi$ (dashed);
(b) the Argand diagram, with masses marked in GeV.}
\end{center}
\end{figure}

The Argand diagram from Ref. \cite {1370} is shown in Fig. 3(b).
It follows a circle closely.
The conclusion is that experimentalists can safely omit
${\rm Re} \, \Pi$ as a first approximation.
However, phase shifts depart significantly from a Breit-Wigner
amplitude of constant width.
At low mass, where $4\pi$ phase space is small, $\Gamma _{total}$ is
small and the phase shift varies rapidly; at high mass it varies
more slowly.
For high quality data, a second pass including ${\rm Re} \, \Pi (s)$
is desirable

Likewise $\eta (1405)$ and $\eta (1475)$ may be fitted as two decay
modes of a single $\eta (1440)$ \cite {1440}.
The $\eta (1475)$ is seen only in $KK^*(890)$, where phase
space rises from threshold near 1385 MeV as momentum cubed in the final
state; the $\eta (1405)$ is seen in $\kappa K$ S-waves and $\eta \pi
\pi$ where phase space changes slowly.

\section {$\sigma$, $\kappa$, $a_0(980)$ and $f_0(980)$}
The $\sigma$ and $\kappa$ poles are well predicted in  both
mass and width by the Roy equations \cite {Caprini} \cite {Descotes},
which are based on $t$- and $u$-channel exchanges.
Exchange of $\rho (770)$ and $K^*(890)$ make strong contributions.
The Julich group of Janssen et al. \cite {Janssen} showed that meson
exchanges account for $f_0(980)$ and $a_0(980)$.
It seems unavoidable that all four states $\sigma$, $\kappa$,
$a_0(980)$ and $f_0(980)$ are strongly driven by
meson exchanges.
They are conspicuous because $q\bar q$ states lie in the mass range
1300--1700 MeV.
There is one feature, however, which is hidden in the meson exchanges
going into the Roy equations.
They impose the Adler zero coming from chiral symmetry breaking.
This is a short-range effect.

Jaffe has proposed \cite {Jaffe2} that $\sigma$, $\kappa$, $a_0(980)$
and $f_0(980)$  are colourless 4-quark states made from a coloured
SU(3) {\bf 3} combination of $qq$ and a ${\bf \bar 3}$ combination of
$\bar q \bar q$.
This naturally leads to a light $\sigma$, an intermediate $\kappa$
and the highest (degenerate) masses for $a_0(980)$ and $f_0(980)$, in
agreement with experiment.
Note, however that meson-meson
configurations lead to a similar spectrum except that the $a_0(980)$
might lie at the $\eta \pi$ threshold.
This does not happen because of
the nearby Adler zero at $s = m^2_\eta - m^2_\pi /2$;
the $a_0(980)$ migrates to the $KK$ threshold because the Adler zero
in this case is distant, at $ s = m^2_K /2$ \cite {nonet}.
Jaffe's  model does not agree
well with the observed decay branching ratio $(\sigma \to KK)/(\sigma
\to \pi \pi)$ near 1 GeV. \cite {Recon}
A serious problem is that, from the width of the $\sigma$ pole, the
$\kappa$ width  is predicted to be $(236 \pm 39)$ MeV, much less than
the latest value: $758 \pm 10(stat) \pm 44 (syst)$ MeV \cite {kappa}.

A further point is that Maiani et al. \cite {Maiani} extend Jaffe's
scheme to $[cq][\bar qq]$ configurations $\bf {6} \otimes \bar {\bf 3}$.
They give a firm prediction for the observation of analogues of
$a_0(980)$ in $c\bar s n\bar n$ with $I=1$ and charges 0, +1 and
+2.
There is no evidence for such states as yet.

Although experimental line-shapes are not affected strongly for broad
resonances, one should not jump to the conclusion that broad resonances
are pure $q\bar q$ states.
The calculations of Oset et al. show that mesonic components are
potentially large.
A related point is what happens to the upper energy combinations
appearing in Eq. (4).
If mesonic components are large, these upper combinations are
moved upwards substantially.
As Jaffe remarks, they become broad and are likely to fall apart,
creating a broad high mass background.
The high mass tail of the $\sigma$ does behave in this sort of way
above 1 GeV, due to coupling to $4\pi$ \cite {1370}; however, the
precise form of this high mass behaviour is poorly known because of
lack of data on $\pi \pi \to 4\pi$.

\section {Applications}
\subsection {$X(3872)$}
It is evident from the width of the cusp in $Re \,\Pi$ of Fig. 1 that
a cusp alone fails to fit the $\sim 3$ MeV width of $X(3872)$
\cite {PDG};
the coupling to $\bar D D^*$ is weaker than that of $f_0(980)$ to $KK$,
but the shape of the dispersion curve is similar.

Several authors conclude that a linear combination of $c\bar c$ and
$\bar D D^*$ is likely in $X(3872)$.
Eichten et al. \cite {Eichten} remark that this explains the low
mass of $X(3872)$ compared to early calculations based purely on
$c\bar c$.
Suzuki \cite {Suzuki} points out that the large production rate of
$X(3872)$ in CDF data from the Tevatron \cite {CDF} requires that it has 
a large wave function at the origin, hence a substantial $c\bar c$ component;
for a pure molecular state the observed production rate is nearly 2
orders of magnitude smaller than CDF observe.
Conversely, the molecular re-arrangements
$\bar D D^* \to \rho J/\Psi$ and $\omega J/\Psi$
account naturally for the weak decays which are observed; here, the
strength of $\bar D D^*$ binding is unimportant.
Bignamini et al. confirm that the CDF production rate is
about 2 orders of magnitude too large to be explained by a molecular
component alone \cite {Bignamini}.
Swanson \cite {SW} also favoures mixing between
$\bar cc$ and a molecular component.

Recently Lee et al. \cite {Lee} have made a detailed fit to existing
data, solving the Bethe-Salpeter equation - equivalent to evaluating
the dispersion integral of Eq.  (2).
The binding energy of $X(3872)$ arises essentially from $\bar D D^*$
loop diagrams.
The magnitude of $\pi $ exchange is known from the width of the decay
$D^* \to D\pi$; other exchanges are modelled.
However, meson exchanges are not the essential source of binding.
They simply need to be attractive, so that $\bar D$ and $D^*$
approach one another.
Both $\bar D_0 D^*_0$ and $D^+ D^-$ channels contribute, though the
$X(3872)$ appears at the lower threshold.
The binding energy is controled sensitively by the form factor.

Kalashnikova and Nefediev conclude that $X(3872)$ has substantial
components of both $\bar cc$ and $\bar D D^*$ \cite {Kalash}.
They point out that Babar data for the ratio of branching ratios
$BR[X(3872)\to \gamma \Psi'(2S)]/ BR[X(3872) \to \gamma J/\Psi ]  = 3.4
\pm 1.4$ \cite {Babarrad} agrees within a factor 2 with estimates for a
$\bar cc$ state \cite {BarnesB}.
The prediction of Swanson for a pure molecule is $\simeq 4 \times
10^{-3}$ \cite {Swanson}.
Also Gutsche et al. conclude that some component of $c\bar c$ is
essential to explain the large rate for $X(3872) \to \gamma \Psi'$
\cite {Gutsche}.
So there seems to be widespread agreement that $X(3872)$ is a linear
combination of $c\bar c$ and $D \bar D^*$.
The precise combination is not yet agreed.

The narrow width of $X(3872)$ arises because its decay modes to
$\pi ^+\pi ^- J/\Psi$, $\omega J/\Psi$ and possibly $\chi (3510)\sigma$
are OZI suppressed, therefore weak.
Its coupling to $\bar D^0 D^{*0}$ over the width of the resonance is
also weak.
However, the coupling to $\bar D D^*$ rises rapidly above threshold
and produces the binding via virtual loop diagrams.
Kalashnikova and Nefediev conclude that the $\bar cc$ state is
attracted to the $\bar D D^*$ threshold.
Ortega et al. \cite {Ortega} reach a similar conclusion that $X(3872)$
must have a large $\bar D D^*$ component.

\subsection {Not all cusps are resonances}
There is a cusp at the $\pi d$ threshold \cite {Measday}, but no
resonance.
The exotic $Z^+(4430)$ of Belle \cite {Z4430} is at the
threshold for $D^*(2007) D_1(2420)$ and has a width close to that of
$D_1(2420)$.
The data can be fitted as a resonance, but can also be
fitted successfully by a non-resonant cusp, see Fig. 6 of \cite {Sync}.
Additionally, Babar do not confirm the existence of the $Z(4430)$.

\subsection {Light Vector Mesons}
Crystal Barrel data in flight, taken together with other data
at lower energies, provide evidence that resonance masses squared for
each spin-parity lie close to straight line trajectories
\cite {Review}.
Updated examples are shown here in Fig. 3.
They resemble Regge trajectories, except they are drawn for one set of
quantum numbers at a time.
There is a striking agreement for all $J^P$, with a common slope
of $1.143 \pm 0.013$ GeV$^2$ for each unit of excitation.
A similar regularity is observed for baryon resonances with
similar slope \cite {Klempt}.
Such regularity agrees with states having a large $q\bar q$ component,
but there is no direct connection with molecules or dynamically
generated states.
\begin{figure}[htb]
\begin{center} \vskip -12mm
\epsfig{file=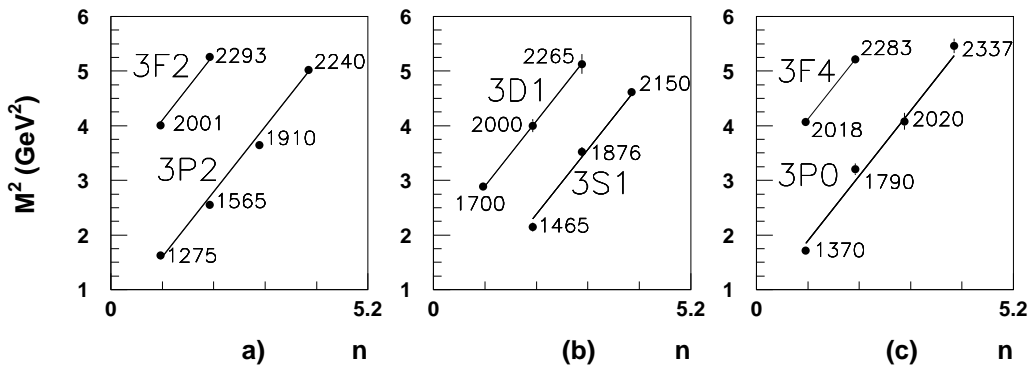,width=16cm}
\vskip -6mm
\caption
{Trajectories of $\bar nn$ resonances: (a) $I=1$, $J^{PC}=1^{--}$,
(b) $I=0$, $J^{PC}=2^{++}$,
(c) $I=0$, $J^{PC}=4^{++}$ and $0^{++}$.}
 \end{center}
\end{figure}

An application of the idea that some resonances mix strongly
with channels to which they decay concerns $\rho (1900)$.
This state lies close to the $N\bar N$ threshold and it is well known
that the $\bar pp \, ^3S_1$ interaction is strongly attractive.
Babar and E687 observe it in decays to $3\pi ^+3\pi ^-$ and
$2(\pi ^+\pi ^-\pi ^0)$ \cite {PDG}.
These are strong decay modes in $N\bar N$ annihilation.
It is natural to interpret  $\rho (1900)$ as the $n=3$ $^3S_1$ $n\bar n$
state mixed with $\bar pp$.
Then other $\rho$ states fall into place as follows:
\newline (ii) $\rho (2000) = ^3D_1,\, n=2$.
It is observed in three sets of data: $\pi ^+\pi ^-$, $\pi \omega$ and
$a_0\omega$.
There are extensive differential cross section and polarisation data
on $\bar pp \to \pi ^+\pi ^-$ from the PS 172 experiment down to a mass
of 1910 MeV (a beam momentum of 360 MeV/c) \cite {172}.
There are further similar data above a beam momentum of 1
GeV/c from an experiment at the Cern PS of Eisenhandler {\it et al}
\cite {Eisenhandler}.
The polarisation  data determine the ratio of decay amplitudes to
$^3D_1$ and $^3S_1$  $\bar pp$ configurations:
$r_{D/S}=g_{\bar pp}(^3D_1)/ g_{\bar pp}(^3S_1)=0.70 \pm 0.32$;
for the low available momentum in $\bar pp$,
this is a rather large $^3D_1$ component.
\newline (iii) $\rho (2150) = ^3S_1,\, n=4$.
It is seen in $\pi ^+ \pi ^-$ data of \cite {172} and
\cite {Eisenhandler} and in  Crystal Barrel data for $a_0(980)\pi$ and
in GAMS and Babar data \cite {PDG}; (the PDG incorrectly lists the 1988
MeV state of Hasan \cite {Hasan} under $\rho (2150)$, but it is the
$\rho (2000)$).
For $\rho (2150)$, $r_{D/S}=-0.05 \pm 0.42$.
\newline (iv) $\rho (2265) = ^3D_1$, $n=3$.
It is observed only in two sets of data, $\pi ^+ \pi ^-$ and in Crystal
Barrel data for $a_2\omega$ and therefore needs confirmation; it has a
large error for $r_{D/S}$.
The $\rho (1700)$, $\rho (2000)$ and $\rho (2265)$ are consistent within
errors with a straight trajectory with the same slope as other states,
see Fig. 2(b).
\newline (v) The $Y(2175)$ \cite {PDG} observed by BES 2 and Babar
in $\phi f_0(980)$ and $K^+K^- f_0(980)$ makes a natural $s\bar s$
partner for $\rho (2000)$.
Note that there is sufficient momentum in the final state to allow a
$^3D_1$ state.

\subsection {$J^P = 2^+$ light mesons}
The $f_2(1565)$ lies at the $\omega \omega$ threshold.
The PDG quotes an average mass of $1562 \pm 13$ MeV, although
Baker et al. \cite {Baker} find a mass of $1598 \pm 11(stat) \pm
9(syst)$ MeV when ${\rm Re} \, \Pi (s)$ is included in the analysis.
This is distinctly lower than the mass of $a_2(1700)$, in the range
1660--1732 MeV.
For almost massless quarks, one expects $f_2$ and $a_2$ masses to be
close.
So $f_2(1565)$  has clearly been attracted to the
$\omega \omega$ threshold.
This requires a molecular component.
The $f_2(1565)$ appears clearly in $\pi \pi$, as observed by several
groups \cite {PDG}.
It should appear in $\rho \rho$ with
$g^2_{\rho \rho} = 3g^2_{\omega \omega}$ by SU(2) symmetry, which
predicts $g(\rho ^0\rho ^0) = - g(\omega \omega)$ because of the
similar masses of light quarks and the close masses of $\rho(770)$
and $\omega (782)$.

The $f_2(1640) \to \omega \omega$ observed by GAMS and VES \cite
{PDG} may be fitted by folding the line-shape of $f_2(1565)$ with
$\omega \omega$ phase space and a reasonable form factor \cite {Baker},
together with the dispersive term ${\rm Re} \, \Pi(s)$.
There is no need for separate $f_2(1565)$ and $f_2(1640)$;
this has confused a number of theoretical predictions of the sequence
of $2^+$ states.

Fig. 2(b) shows trajectories for $2^{++}$ states, including those above
the $\bar pp$ threshold from Crystal Barrel data in flight, using
trajectories with a slope of 1.14 GeV$^2$.
The PDG makes a number of serious errors in reporting the Crystal
Barrel publications.
Those mistakes distort conclusions to be drawn from the data.
It lists $f_2(2240)$ under $f_2(2300)$, which is observed in $\phi
\phi$ and $KK$ by all other groups.
The $f_2(2300)$ is naturally interpreted as an $s\bar s$ state.
Both $f_2(2240)$ and $f_2(2293)$  are observed in a combined
analysis of ten sets of data: four sets  of PS172 and Eisenhandler et
al., together with Crystal Barrel data for $\eta \pi ^0 \pi ^0$, $\eta
'\pi ^0 \pi ^0$, $\eta \eta \eta$, $\pi ^0\pi ^0$, $\eta \eta$ and
$\eta \eta '$.
The data from the last 3 channels are fitted to a linear
combination $\cos \phi \, |n\bar n> + \sin \phi \, |s\bar s>$ and the
mixing angle is determined to be $\phi = 7.5^\circ$ for $f_0(2240)$ and
$\phi = -14.8^\circ$ for $f_2(2293)$ \cite {mixing}.
So the $f_2(2240)$ is certainly not an $s\bar s$ state.
From polarisation data, the $f_2(2240)$ is dominantly $^3P_2$ with
$r_{F/P} = 0.46 \pm 0.09$ (defined like $r_{D/S}$) and the $f_2(2300)$
is largely $^3F_2$ with $r_{F/P} = -2.2 \pm 0.6$.
The PDG fails to list the $f_2(2293)$ at all, despite many prompts
over a 6 year period.
It is observed in 5 channels: $\pi \pi$, $\eta \eta$, $\eta \eta '$,
$f_2\eta$ and $a_2(1320)\pi$.

A further comment is that the $f_2(2150)$ is
conspicuous by its absence from Crystal Barrel data in flight.
All $s\bar s$ states such as $f_2(1525)$ are produced very
weakly in $\bar pp$ interactions.
The $f_2(2150)$ is observed mostly in $K\bar K$ and $\eta \eta$
channels.
It is therefore naturally interpreted as an $s\bar s$ state, the
partner of $f_2(1905)$.

An important systematic observation is that $\bar pp$ states tend to
decay with the same $L$ as the initial $\bar pp$ state.
There is a simple explanation, namely good overlap of the initial
and final states in impact parameter.
This observation may be useful to those calculating decays, hence
mesonic contributions to eqns. (1)--(4).

\subsection {Light $0^+$ mesons}
There is extensive evidence for a radial excitation of $f_0(1370)$
at 1790 MeV.
The facts run as follows.
In BES 2 data for $J/\Psi \to \omega K^+K^-$, there is a clear
$f_0(1710) \to KK$ \cite {WKK}.
In high statistics data for $J/\Psi \to \omega \pi ^+\pi ^-$, \cite
{WPP} there is no visible $f_0(1710)$, setting a limit on branching
ratios:  $BR(f_0(1710) \to \pi \pi )/BR (f_0(1710) \to KK) <0.11$ with
95\% confidence.
Thirdly, in $J/\Psi \to \phi \pi ^+\pi ^-$, there is a
$\pi \pi$ peak requiring an {\it additional} $f_0(1790)$ decaying to
$\pi \pi$ but weakly to $KK$ \cite {phipp}.
There is ample independent evidence for it in $J/\Psi \to \gamma 4\pi$
\cite {Mark3} \cite {Bai} \cite {g4pi} and
$\bar pp \to \eta \eta \pi ^0$ in flight \cite {eepi}.
BES 2 also report an $\omega \phi$ peak of
95 events at 1812 MeV; $J^P = 0^+$ is favoured \cite {omegaphi}.
It is confirmed by VES data at the Hadron09 conference \cite
{Ivashin}. The BES data may be fitted well with the $f_0(1790)$
line-shape folded with $\omega \phi$ phase space and a form factor
$\exp -3k^2_{\omega \phi}$.
There is some scatter on Fig. 4(c) of masses about the line of
standard slope; this may well be because masses of $0^+$ states
tend to be the most difficult to deterine due to isotropic angular
distributions.

A comment is needed on $f_2(1810)$ of the Particle Data Tables.
It does not fit in naturally in Fig. 2(b).
The spin analysis of the GAMS group \cite {Alde} finds a very marginal
difference between spin 0 and spin 2.
It rests on a fine distinction in the angular
distribution depending strongly on experimental acceptance; however, no
Monte Carlo of the acceptance is shown.
With the benefit of hindsight, it seems possible that this was in fact
the first observation of $f_0(1790)$.

Summarising subsections 5.3 to 5.5, there is strong evidence in
Figs. 4(a)-(c) that resonances lie close to straight trajectories
as a function of mass squared.
These may be redrawn as Regge trajectories for $1^-$, $2^+$, $3^-$,
etc., see Fig. 56 of \cite {Review}.
Regge trajectories are naturally explained by a flux tube joining $q$
and $\bar q$; the energy stored in the flux tube explains the linear
relation between $J$ and mass squared.
There is no clear reason why molecules or dynamically generated states
should follow such trajectories.

\subsection {Glueballs}
Morningstar and Peardon \cite {glueball}
predict glueball masses in the quenched approximation where $q\bar q$
are omitted.
When mixing with $q\bar q$ is included, mixing is likely to lower
glueball masses.

\section {Remarks on further experiments}
Further progress towards a complete spectroscopy of light mesons and
baryons is important for an understanding of confinement - one of the
key phase transitions in physics.
Progress is possible by measuring transverse polarisation in formation
processes.
Consider $\bar pp$ as an example.
The high spin states appear clearly as peaks, e.g. $f_4(2050)$ and
$f_4(2300)$.
These serve as interferometers for lower states.
However, differential cross sections measure only real parts of
interferences.
This leaves the door open to two-fold ambiguities in relative phases
and large errors if resonances happen to be orthogonal.
A measurement of transverse polarisation normal to the plane of
scattering measures $Tr <A^*\sigma _yA>$,
where $A$ is the amplitude.
This measures the imaginary part of interferences.
What appears to be less well known is that transverse polarisation in
the plane of scattering gives additional information for three and
four-body final states with a decay plane different to the plane
defined by the beam and initial state polarisation.
This depends on $Tr <A^*\sigma _xA>$, and measures the real parts of
exactly the same interferences as appear from the $\sigma_y$ operator.
Longitudinal polarisation depends only on differences of two
intensities and is less useful.

An example of a simple experiment which would pay a rich dividend
is to measure such polarisations with the Crystal Barrel detector
at the forthcoming GSI $\bar p$ source, over the same mass range as
used at LEAR.
An extracted beam with these momenta will be available at the FLAIR
ring.
Such measurements could indeed have been made at LEAR if it had not
been sacrificed to the funding of the LHC.
The present situation is that the amplitudes for $I=0$, $C=+1$ states
are unique for all expected $J^P$.
For $I=1$, $C=-1$, they are nearly complete, but there are some
weaknesses for low spin states, notably $^3S_1$ which leads to a flat
decay angular distribution.
For $I=1$, $C=+1$ there is a two-fold ambiguity for $\eta \pi$ final
states and crucial $J^P = 0^+$ states are missing.
For $I=0$, $C=-1$ there are many missing states.

The measurements required are to (i) $\eta \pi$ and
$\eta \eta \pi$ ($I=1$, $C = +1$), (ii) $\omega \pi$ ($I=1$, $C=-1$),
(iii) $\omega \eta$ and $\omega \eta \pi^0$ ($I=0$, $C=-1$).
A measurement of $\eta \pi ^0 \pi ^0$ would cross-check the existing
solution and provide information on interferences between singlet
and triplet $\bar pp$ states.
As well as locating missing states, this would build up a clear
picture of the many decays observable, for comparision with
meson exchange processes.
All of these channels can be measured simultaneously with the existing
Crystal Barrel detector.

A Monte Carlo simulation of results extrapolated from existing analyses
predicts a unique set of amplitudes for all quantum numbers.
Data are required from 2 GeV/c down to the lowest possible momentum
$\sim 360$ MeV/c.
Seven of the nine momenta studied at LEAR were run in 3 months of beam
time, so it is not a long experiment, nor does it demand beam
intensities above $5 \times 10^4 \bar p/s$.
A Monte Carlo study shows that backgrounds from heavy nuclei in the
polarised target (and its cryostat) should be at or below an average
level of 10\%;
this is comparable with cross-talk between final states and is easily
measured from a dummy target.

Baryon spectrocopy would also benefit from similar $\pi ^\pm p$
polarisation measurements  in inelastic channels.
A transverse magnetic field is required for compatibility with the
polarised target.
Rates are enormous, so running time is governed by down-time
required for polarising the target and changing momenta.
Data at 30 MeV steps of mass appear sufficient, except close to 2-body
thresholds such as $\omega N$.

\section {Conclusions}
The objective of this paper has been to make a case for what appears
logically necessary, namely that both quark combinations at short
range and decay channels at large range contribute to the
eigenstates.
The $X(3872)$ is a prime example of mixing between $c\bar c$ and
meson-meson in the form of $\bar D D^*$.

In view of the calculations of Oset et al. and Barnes and Swanson, it
seems likely that many resonances contain large mesonic contributions.
The straight trajectories of Figs. 2(a)-(c) are naturally explained by
Regge phenomenology; molecules and dynamically generated states provide
no obvious explanation of these trajectories.
The large mass shift between $f_2(1565)$ and $a_2(1700)$ indicates
meson mixing into $f_2(1565)$, lowering the eigenvalue in analogy
with the covalent bond in chemistry.

The data on $a_0(980)$, $f_0(980)$, $f_2(1565)$ and $\rho(1900)$
fit naturally into this picture.
There must be a large mesonic contribution to the nonet of $\sigma$,
$\kappa$, $a_0(980)$ and $f_0(980)$, but there could be a modest
diquark component as well.
It is likely that there will be a small
$q\bar q$ component, but this is suppressed by the $L=1$ centrifugal
barrier for $^3P_0$ combinations.

Experimentalists must take care to fit the $s$-dependence of the
numerator of Breit-Wigner resonances due to phase space,
e.g. $f_0(1370) \to 2\pi$ has a very different line-shape to
$f_0(1370) \to 4\pi$.
The denominator may be fitted as a first approximation with a
Breit-Wigner resonances of constant width; however, for high quality
data, the effect of the dispersive component in the real part of the
denominator matters.
For sharp thresholds, e.g. $f_0 \to KK$, the Flatt\' e formula is an
approximation; with high quality data, the correction due to the cusp
in Re $\Pi(s)$ is important, but requires precise information on
experimental resolution.

Further experiments on transverse polarisation in inelastic processes
are needed and appear to be practicable without large cost.

\end{document}